\documentclass[10pt,a4papper]{article}
\usepackage{indentfirst}
\usepackage[english]{babel}
\usepackage{geometry} 
\usepackage{graphicx}
\begin{document}

\title{\textbf{Low energy process $e^{+}e^{-} \rightarrow K^{+} K^{-}$ in the extended Nambu-Jona-Lasinio model}}
\author{M. K. Volkov$^{1}$\footnote{volkov@theor.jinr.ru}, K. Nurlan$^{1,2,3}$\footnote{nurlan.qanat@mail.ru}, A. A. Pivovarov$^{1}$\footnote{tex$\_$k@mail.ru}\\
\small
\emph{$^{1}$ Bogoliubov Laboratory of Theoretical Physics, Joint Institute for Nuclear Research, Dubna, 141980, Russia}\\
\small
\emph{$^{2}$ Institute of Nuclear Physics, Almaty, 050032, Republic of Kazakstan}\\
\small
\emph{$^{3}$ Dubna State University, Dubna, 141980, Russia}}
\date{}
\maketitle
\small

\begin{abstract}

In the extended Nambu-Jona-Lasinio model, the process of production of charged kaon pair on colliding electron-positron beams
in the energy interval 1 - 1.7 GeV is described. In studying of this process, we take into account the contact terms when the pairs $K^{+}$ $K^{-}$
are generated by an intermediate photon, and processes with intermediate vector mesons $\rho$, $\omega$ and $\phi$ in the ground and in the first
radially excited states. The results obtained here are compared with experimental data recently
received in Novosibirsk and in Stanford.

\end{abstract}
\large
\section{Introduction}

Recently, great attention has been paid to the experimental study of the production of charged kaon
pairs in colliding electron-positron beams in the energy range 1 - 2 GeV at the Budker Institute of Nuclear Research (Novosibirisk) \cite{Achasov:2000am, Achasov:2007kg, Akhmetshin:2008gz, Achasov:2016lbc, Kozyrev:2017agm} and in the SLAC National Accelerator Laboratory (Stanford) \cite{Lees:2013gzt}. For a theoretical description of this process in the low-energy region, unfortunately, we can not use the QCD perturbation theory, but here we can successfully apply various phenomenological models based, as a rule, on the chiral symmetry of strong interactions. Among them, we should like to note the chiral perturbations theory \cite{Gasser:1983yg, Gasser:1984ux, Ecker:1988te}, since this model is preferable only at lower energies not grafting the mass of the $\rho$ meson. At present, a number of models based on the above chiral perturbation theory have appeared, which consider the extended version, for example, the extended vector-meson-dominance model \cite{Ivashyn:2006gf}. In this model, the experimental results of SND \cite{Achasov:2000am} have recently been described, but to describe the radially excited intermediate states in this model it was necessary to use a number of additional arbitrary parameters.

A well-known model of this type is also the Nambu-Jona-Lasinio (NJL) model. In addition to the standard NJL model \cite{Nambu:1961tp, Eguchi:1976iz, Kikkawa:1976fe, Volkov:1982zx, Ebert:1982pk, Volkov:1984kq, Volkov:1986zb, Ebert:1985kz, Vogl:1991qt, Klevansky:1992qe, Volkov:1993jw, Ebert:1994mf, Volkov:2006vq} that is used to describe mesons in ground states and their interactions at low energies, there is also an extended version of the NJL model \cite{Volkov:2006vq, Volkov:1996fk, Volkov:1996br, Volkov:1997dd, Volkov:1999yi}, which allows one to describe not only the ground states but also the first radially excited meson states, without violating $U(3) \times U(3)$ chiral symmetry. This is achieved by using the simplest form factors, that have the form of first-order polynomials in the square of the quark momenta. The form factor has the form $ F (\vec{k}^{2}) = c f\left(\vec{k}^{2}\right) $, $f\left(\vec{k}^{2}\right) = 1 + d \vec{k}^{2}$, here $\vec{k}^{2}$  is the quark momentum. The constant $c$ only affects the radially excited meson masses. The slope parameter $d$ is chosen so that the radially-excited state does not influenced quark condensate and, hence, the values of the constituting quark masses. These masses are the main parameters of the NJL model. In this case, the quark loop having the form of a tadpole containing the form factor $f\left(\vec{k}^{2}\right)$ should be equated to zero. After the introduction of the first radially excited meson states, nondiagonal terms appear in the NJL model in the free Lagrangian. These terms correspond to exchanges between the meson states with and without form factors. The free Lagrangian is diagonalized by introducting mixing angles \cite{Volkov:1996fk, Volkov:1999yi, Volkov:2016umo, Volkov:2017arr}. After diagonalizing the free Lagrangian, we can describe various meson interactions in both the ground and first radially excited states without introducing any arbitrary parameters. In particular for the description of $e^{+}e^{-} \rightarrow K^{+} K^{-}$ at energy 1-1.7 GeV intermediate $\rho$, $\omega$, $\phi$, $\rho(1450)$, $\omega(1420)$, $\phi(1680)$ vector states are considered.

In recent years, by using the extended NIL model, the following processes have been described:
$e^{+}e^{-} \rightarrow \left[\gamma(\pi, \pi(1300), \eta, \eta'(958), \eta(1295), \eta(1475)); \pi(\pi, \pi(1300)); 2\pi(\eta,\eta'(958)); \pi\omega \right]$ \cite{Volkov:2016umo, Volkov:2017arr},
$e^{+}e^{-} \rightarrow \left[\eta(\phi(1020),\rho); K^{\pm} (K^{*\mp}(892))\right]$ \cite{Volkov:2016zdw, Volkov:2017cmv}.
The process $e^{+}e^{-} \rightarrow K^{+} K^{-}$, considered in the present paper, naturally supplements the previously considered series of similar processes. In the review \cite{Volkov:2017arr}, it was shown that the extended
NJL model makes it possible to successfully describe not only the numerous meson production processes in colliding electron-positron beams but also many of the main decays of tau lepton.

\section{Effective quark-meson Lagrangian \\}

In the extended NJL model, the quark-meson interaction Lagrangian for pseudoscalar $K^{\pm}$,
vector $\rho, \omega, \phi$ mesons in the ground and first radially excited states takes the form:

\begin{eqnarray}
{\cal L}_{K} = \bar{q}\{i\gamma^{5}\sum_{j = \pm}\lambda_{j}(a_{K}K^{j} + b_{K}K'^{j})\}q,
\end{eqnarray}

\begin{eqnarray}
{\cal L}_{\rho} = \bar{q}\{\frac{1}{2}\gamma^{\mu}\lambda_{\rho}(a_{\rho}\rho_{\mu} + b_{\rho}\rho'_{\mu})\}q,
\end{eqnarray}

\begin{eqnarray}
{\cal L}_{\omega} = \bar{q}\{ \frac{1}{2}\gamma^{\mu}\lambda_{\omega}(a_{\omega}\omega_{\mu} + b_{\omega}\omega'_{\mu})\}q,
\end{eqnarray}

\begin{eqnarray}
{\cal L}_{\phi} = \bar{q}\{\frac{1}{2}\gamma^{\mu}\lambda_{\phi}(a_{\phi}\phi_{\mu} + b_{\phi}\phi'_{\mu}) \}q,
\end{eqnarray}

where $q$ and $\bar{q}$ are the u-, d- and s- constituent quark fields with masses $m_{u} = m_{d} = 280$ MeV,
$m_{s} = 420$ MeV \cite{Volkov:2016umo, Volkov:2017arr, Volkov:2001ns}, $K^{\pm}$, $\rho$, $\omega$ and $\phi$ are
the pseudoscalar and vector mesons, the excited states are marked with prime,

\begin{displaymath}
a_{a} = \frac{1}{\sin(2\theta_{a}^{0})}\left[g_{a}\sin(\theta_{a} + \theta_{a}^{0}) +
g_{a}^{'}f_{a}(\vec{k}^{2})\sin(\theta_{a} - \theta_{a}^{0})\right],
\end{displaymath}
\begin{equation}
\label{Coefficients}
b_{a} = \frac{-1}{\sin(2\theta_{a}^{0})}\left[g_{a}\cos(\theta_{a} + \theta_{a}^{0}) +
g_{a}^{'}f_{a}(\vec{k}^{2})\cos(\theta_{a} - \theta_{a}^{0})\right],
\end{equation}
$f\left(\vec{k}^{2}\right)$ is the form factor, $\theta_{a}$ and $\theta_{a}^{0}$ are the mixing angles for the mesons in the ground and excited states \cite{Volkov:1996fk, Volkov:2016umo, Volkov:2017arr}.
 The slope parameters and mixing angles are

\begin{displaymath}
d_{uu} = -1.784 \textrm{GeV}^{-2},  \quad d_{ss} = -1.737 \textrm{GeV}^{-2},
\end{displaymath}
\begin{equation}
\begin{array}{cccc}
\theta_{K} = 58.11^{\circ},     & \theta_{\rho} = \theta_{\omega} = 81.8^{\circ},        & \theta_{\phi} = 68.4^{\circ},\\
\theta_{K}^{0} = 55.52^{\circ},& \theta_{\rho}^{0} = \theta_{\omega}^{0} = 61.5^{\circ},& \theta_{\phi}^{0} = 57.13^{\circ}.
\end{array}
\end{equation}

The matrices

\begin{displaymath}
\lambda_{\rho} = \left(\begin{array}{ccc}
1 & 0  & 0\\
0 & -1 & 0\\
0 & 0  & 0
\end{array}\right), \quad
\lambda_{\omega} = \left(\begin{array}{ccc}
1 & 0  & 0\\
0 & 1  & 0\\
0 & 0  & 0
\end{array}\right), \quad
\lambda_{\phi} = -\sqrt{2} \left(\begin{array}{ccc}
0 & 0  & 0\\
0 & 0  & 0\\
0 & 0  & 1
\end{array}\right), \quad
\end{displaymath}

\begin{displaymath}
\lambda_{+} = \sqrt{2} \left(\begin{array}{ccc}
0 & 0 & 1\\
0 & 0 & 0\\
0 & 0 & 0
\end{array}\right), \quad
\lambda_{-} = \sqrt{2} \left(\begin{array}{ccc}
0 & 0 & 0\\
0 & 0 & 0\\
1 & 0 & 0
\end{array}\right),
\end{displaymath}

The coupling constants:

\begin{displaymath}
g_{K} = \left(\frac{4}{Z_{K}}I_{2}(m_{u},m_{s})\right)^{-1/2} \approx 3.77,
\quad g_{K}^{'} = \left(4I_{2}^{f_{us}^{2}}(m_{u},m_{s})\right)^{-1/2} \approx 4.69,
\end{displaymath}
\begin{displaymath}
g_{\rho} = g_{\omega} = \left(\frac{2}{3}I_{2}(m_{u},m_{u})\right)^{-1/2} \approx 6.14,
\quad g_{\rho}^{'} = g_{\omega}^{'} = \left(\frac{2}{3}I_{2}^{f_{uu}^{2}}(m_{u},m_{u})\right)^{-1/2} \approx 9.87,
\end{displaymath}
\begin{displaymath}
g_{\phi} = \left(\frac{2}{3}I_{2}(m_{s},m_{s})\right)^{-1/2} \approx 7.5,
\quad g_{\phi}^{'} = \left(\frac{2}{3}I_{2}^{f_{ss}^{2}}(m_{s},m_{s})\right)^{-1/2} \approx 13.19,
\end{displaymath}
where

\begin{equation}
Z_{K} = \left(1 - \frac{3}{2}\frac{(m_{u} + m_{s})^{2}}{M^{2}_{K_{1}}}\right)^{-1} \approx 1.83,
\end{equation}
$Z_{K}$ is the factor corresponding to the $K - K_{1}$ transitions,
 $M_{K_{1}} = 1272$ MeV \cite{Patrignani:2016xqp} is the mass
of the axial-vector  $K_{1}$ meson, and the integral $I_{2}$ has the following form:

\begin{equation}
I_{2}^{f^{n}}(m_{1}, m_{2}) =
-i\frac{N_{c}}{(2\pi)^{4}}\int\frac{f^{n}(\vec{k}^{2})}{(m_{1}^{2} - k^2)(m_{2}^{2} - k^2)}\theta(\Lambda_{3}^{2} - \vec{k}^2)
\mathrm{d}^{4}k,
\end{equation}
$\Lambda_{3} = 1.03$ GeV is the cut-off parameter \cite{Volkov:1996fk,Volkov:1999yi}.

All these parameters were defined earlier and are standard for the extended NJL model.

\section{The amplitude of the process $e^{+}e^{-} \rightarrow K^{+} K^{-}$}
\begin{figure}[h]
\center{\includegraphics[scale = 0.6]{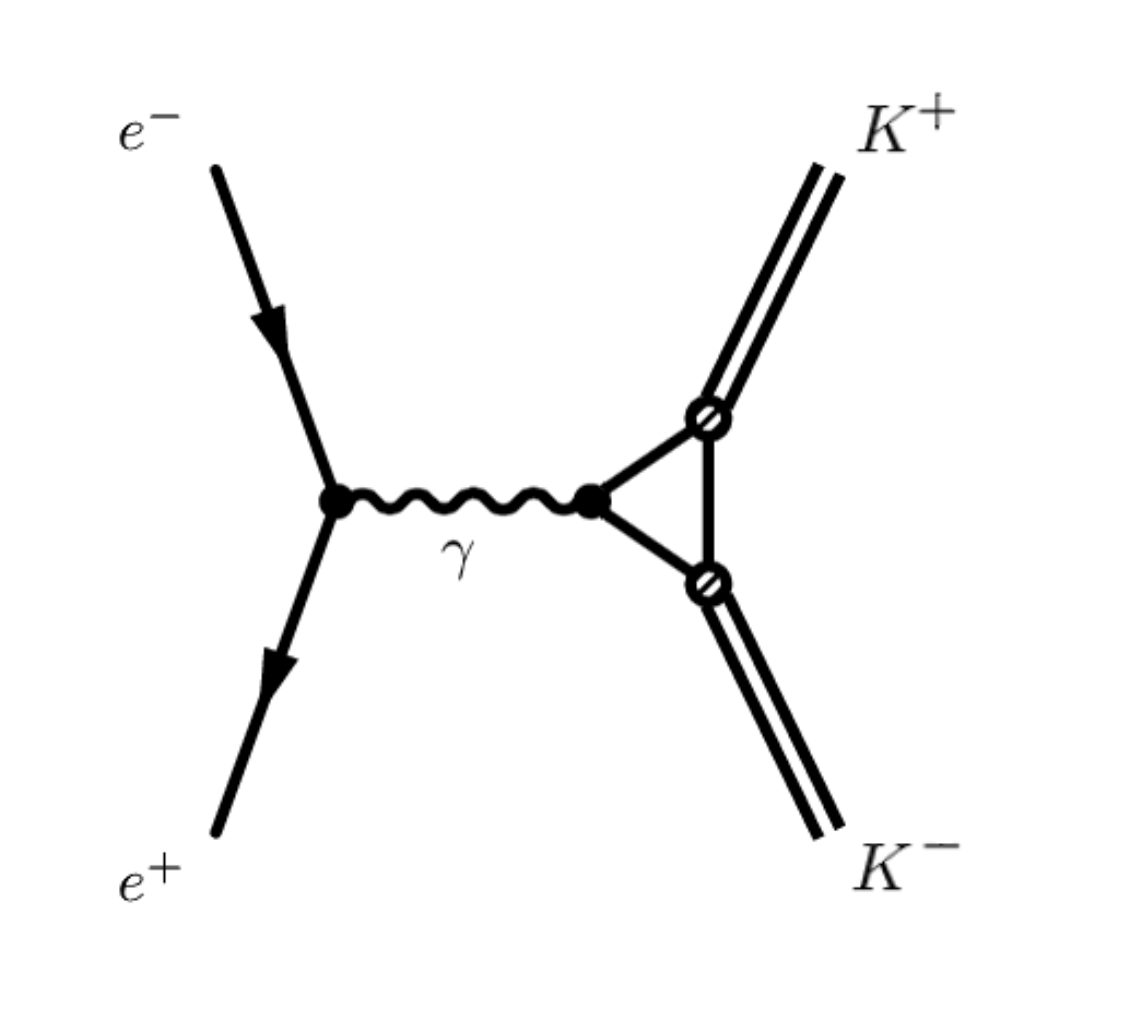}}
\caption{The Feynman diagram with photon exchange(contact diagram).}
\label{Contact}
\end{figure}

\begin{figure}[h]
\center{\includegraphics[scale = 0.8]{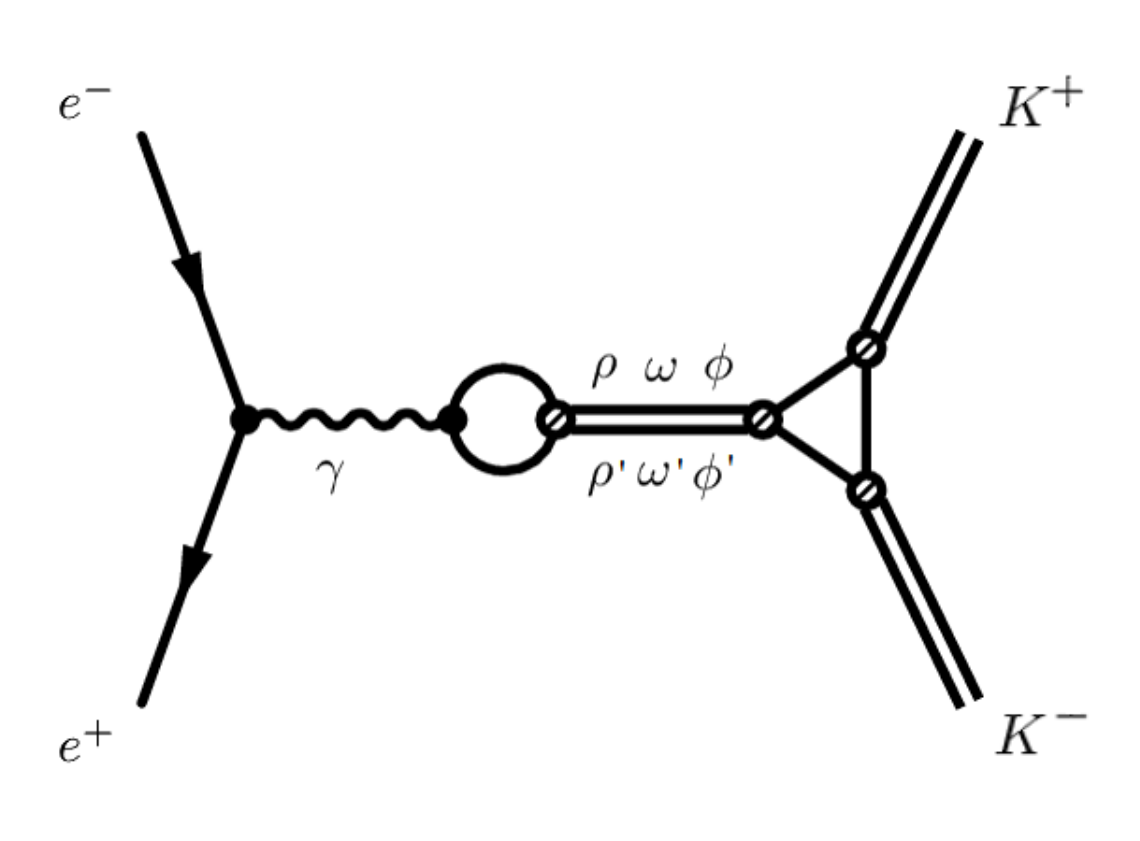}}
\caption{Feynman diagram(s) with intermediate vector meson exchange.}
\label{Intermediate}
\end{figure}

The diagrams of the process $e^{+}e^{-} \rightarrow K^{+} K^{-}$ are shown in Figs.\ref{Contact},\ref{Intermediate}.


The process $e^{+}e^{-} \rightarrow K^{+} K^{-}$ contains contributions of following amplitudes:

\begin{equation}
T = \frac{16 \pi \alpha_{em}}{s} l^{\mu} \left\{B_{(\gamma)} + B_{(\rho + \rho')} + B_{(\omega + \omega')} +
e^{i\pi}B_{(\phi + \phi')}\right\}_{\mu\nu}(p_{K^{+}}-p_{K^{-}})^{\nu},
\end{equation}
where $s = (p(e^{-}) + p(e^{+}))^2$, $l^{\mu} = \bar{e}\gamma^{\mu}e$ is the lepton current. Unfortunately, the NJL model can not describe a relative phase between different states. Thus, we take a phase from $e^{+} e^{-}$ annihilation experiments \cite{Achasov:2000am, Achasov:2007kg} ($e^{i\pi}$ factor in the $\phi$ mesons). The contribution of the contact diagram is

\begin{equation}
B_{(\gamma)\mu\nu} =  g_{\mu\nu}I^{\gamma KK}_{2},
\end{equation}
where $\gamma = 1$.
The sum of $\rho$ and $\rho'$ meson contributions reads

\begin{equation}
B_{(\rho + \rho')\mu\nu} = \frac{1}{2}
\left[\frac{C_{\rho}}{g_{\rho}}\frac{g_{\mu\nu}s - p_{\mu}p_{\nu}}{M^{2}_{\rho} - s - i\sqrt{s}\Gamma_{\rho}(s)} I^{\rho KK}_{2} + \frac{C_{\rho'}}{g_{\rho}}\frac{g_{\mu\nu}s - p_{\mu}p_{\nu}}{M^{2}_{\rho'} - s - i\sqrt{s}\Gamma_{\rho'}(s)} I^{\rho' KK}_{2}\right].
\end{equation}
The contribution of the diagrams with the intermediate $\omega$, $\omega'$ meson is

\begin{equation}
{B}_{(\omega + \omega')\mu\nu} = \frac{1}{6}
\left[\frac{C_{\omega}}{g_{\omega}}\frac{g_{\mu\nu}s - p_{\mu}p_{\nu}}{M^{2}_{\omega} - s - i\sqrt{s}\Gamma_{\omega}(s)} I^{\omega KK}_{2}
+\frac{C_{\omega'}}{g_{\omega}}\frac{g_{\mu\nu}s - p_{\mu}p_{\nu}}{M^{2}_{\omega'} - s - i\sqrt{s}\Gamma_{\omega'}(s)} I^{\omega'KK}_{2}\right].
\end{equation}
The contribution of the diagrams with the intermediate $\phi$, $\phi'$ meson is

\begin{equation}
{B}_{(\phi + \phi')\mu\nu} = \frac{1}{3}
\left[\frac{C_{\phi}}{g_{\phi}}\frac{g_{\mu\nu}s - p_{\mu}p_{\nu}}{M^{2}_{\phi} - s - i\sqrt{s}\Gamma_{\phi}(s)} I^{\phi KK}_{2}+ \frac{C_{\phi'}}{g_{\phi}}\frac{g_{\mu\nu}s - p_{\mu}p_{\nu}}{M^{2}_{\phi'} - s - i\sqrt{s}\Gamma_{\phi'}(s)} I^{\phi'KK}_{2}\right].
\end{equation}
$M_{\rho} = 775$ MeV, $M_{\rho'} = 1465$ MeV, $M_{\omega} = 783$ MeV, $M_{\omega'} = 1420$ MeV, $M_{\phi} = 1019$ MeV, $M_{\phi'} = 1680$ MeV are the masses of the intermediate vector mesons \cite{Patrignani:2016xqp}. Here, instead of the constant decay width, we used $\Gamma(s)$ like in \cite{Lees:2013gzt}:

\begin{equation}
\Gamma_{V}(s) = \Gamma_{V}\frac{s}{M^{2}_{V}}{\left(\frac{\beta(s, M_{K})}{\beta(M^{2}_{V}, M_{K})}\right)}^{3},
\end{equation}
where $\beta(s, M_{K})=\sqrt{1-{4M_{K}}^2/s}$.

The numerical coefficients $C_{a}$ are obtained from the quark loops in the transitions of the photon into the intermediate vector mesons:

\begin{equation}
\label{gamma-in-meson}
C_{a} = \frac{1}{\sin\left(2\theta_{a}^{0}\right)}\left[\sin\left(\theta_{a} + \theta_{a}^{0}\right) +
R_{V}\sin\left(\theta_{a} - \theta_{a}^{0}\right)\right],
\end{equation}

\begin{displaymath}
R_{V} = \frac{I_{2}^{f}(m_{1},m_{2})}{\sqrt{I_{2}(m_{1},m_{2})I_{2}^{f^{2}}(m_{1},m_{2})}},
\end{displaymath}
where $m_{1}$ and $m_{2}$ are the masses of the u-quarks or the s-quarks depending on the quark structure of the intermediate vector meson.
The integrals

\begin{equation}
I_2^{abc}(m_{u}, m_{s}) = -i\frac{N_{c}}{(2\pi)^{4}}\int\frac{a(\vec{k}^{2})b(\vec{k}^{2})c(\vec{k}^{2})}{(m_{u}^{2} - k^2)(m_{s}^{2} - k^2)}
\theta(\Lambda_{3}^{2} - \vec{k}^2) \mathrm{d}^{4}k,
\end{equation}
are obtained from the quark triangles, $a(\vec{k}^{2})$, $b(\vec{k}^{2})$ and $c(\vec{k}^{2})$ are the coefficients defined in (\ref{Coefficients}).

\section{Numerical estimations}
The cross section of the process $e^{+}e^{-} \rightarrow K^{+} K^{-}$ as a function of energy is shown in Figs 3 and 4.
\begin{figure}[h]
\center{\includegraphics[scale = 0.5]{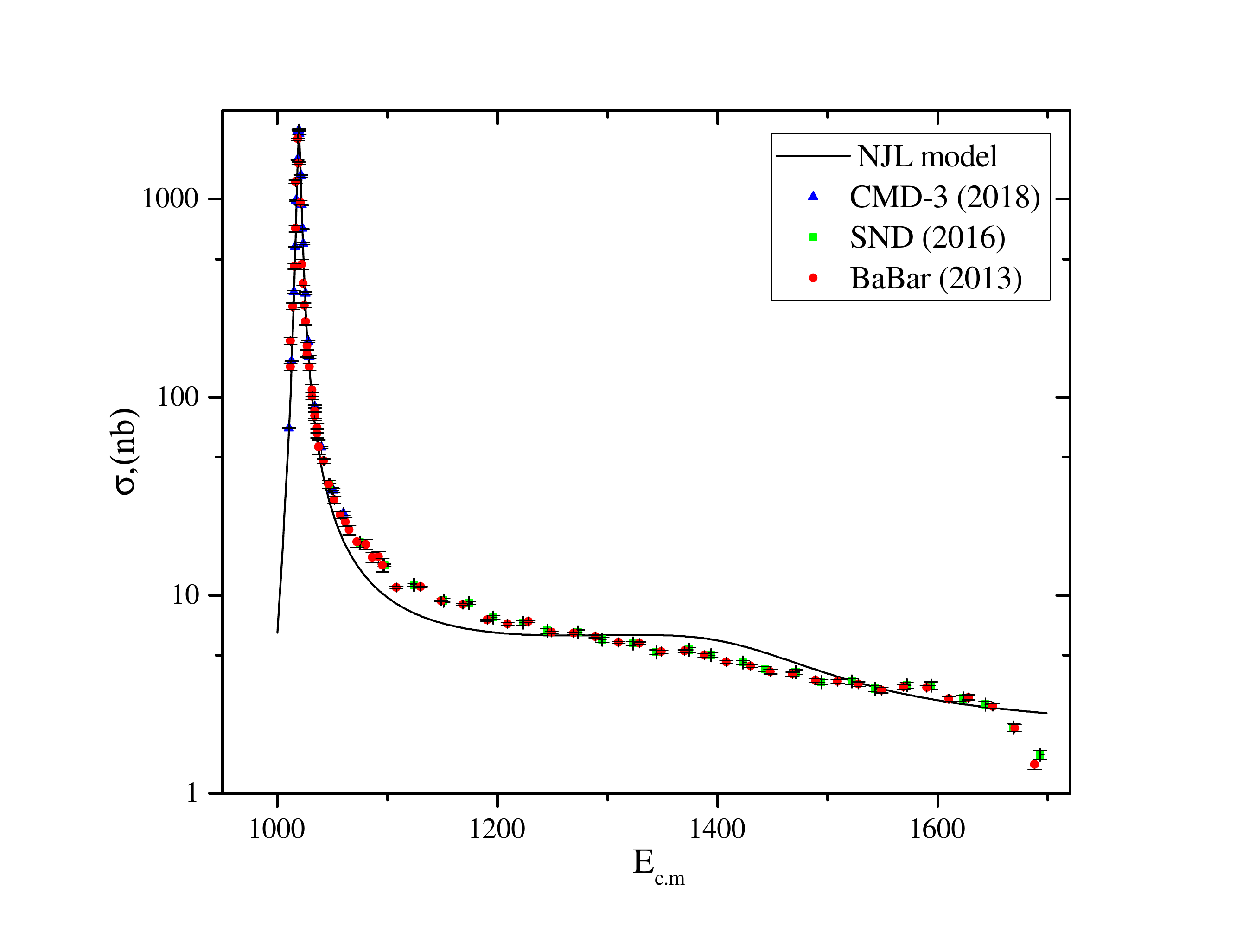}}
\caption{The figure in the logarithmic scale shows the theoretical predictions of the NJL model for the process $e^{+}e^{-} \rightarrow K^{+} K^{-}$ with the account of six intermediate meson states.
The experimental data \cite{ Achasov:2016lbc, Kozyrev:2017agm, Lees:2013gzt} are shown as separate points.}
\label{CrossSection1}
\end{figure}

\begin{figure}[h!]
\center{\includegraphics[scale = 0.4]{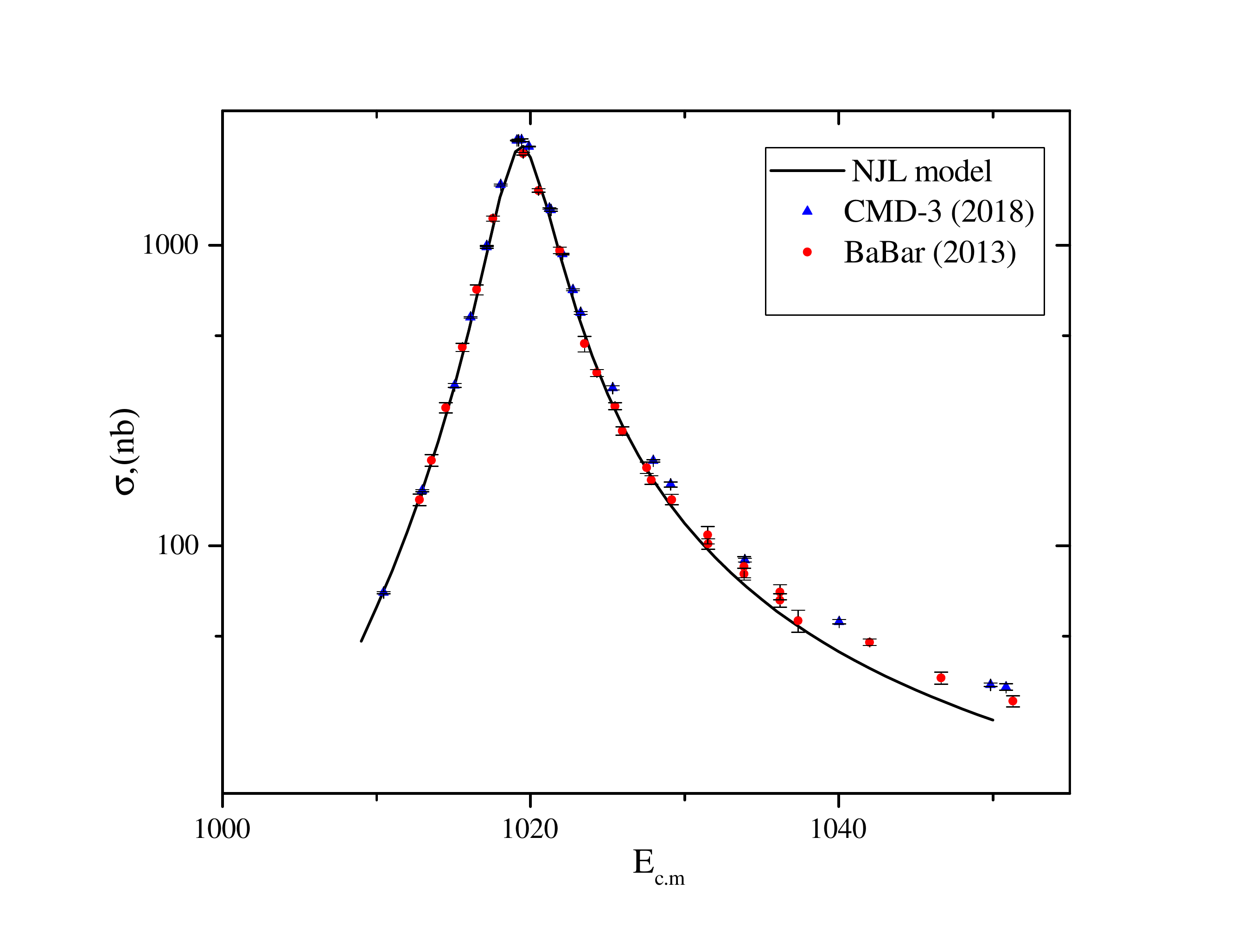}}
\caption{The resonance region of the $\phi$ meson in the process $e^{+}e^{-} \rightarrow K^{+} K^{-}$.}
\label{CrossSection2}
\end{figure}

The results are in satisfactory agreement with the experimental data of SND, CMD-3 \cite{ Achasov:2016lbc, Kozyrev:2017agm} and the BaBar Collaboration
\cite{Lees:2013gzt} in the energy range 1 - 1.6 GeV (see fig.3,4). It is also interesting to compare our results with the results obtained in other fenomenological models \cite{Ivashyn:2006gf, Bruch:2004py}, in particular, the vector-meson-dominance model, which is based on the chiral perturbation theory mentioned in the introduction. These results together with our results are summarized in Table 1.

\begin{table}[h]
	\caption{The comparison of the absolute values of the factors in the numerators of the Breit-Wigner propagators, that describes the transition $\gamma \rightarrow V \rightarrow K^{+} K^{-}$}
	\label{Tab}
	\begin{center}
		\begin{tabular}{|c|c|c|c|}
			\hline
							& our result & \cite{Ivashyn:2006gf} & \cite{Bruch:2004py}      \\
			\hline
			$N_{\rho}$      &  0.44   &  0.598  &  0.598   	 \\
			$N_{\omega}$    &  0.147  &  0.171  &  0.199 	 \\
			$N_{\phi}$		&  0.34   &  0.283  &  0.339	 \\
			$N_{\rho'}$ 	&  0.066  &  0.056  &  0.056 	 \\
			$N_{\omega'}$	&  0.022  &  0.016  &  0.019     \\
			$N_{\phi'}$ 	&  0.0005 &  0.005  &  0.006     \\
			\hline
		\end{tabular}
	\end{center}
\end{table}

In the table, the absolute values of the numerators of the Breit-Wigner propagators of the form factors are given. In reference \cite{Ivashyn:2006gf} in table 4 a comparison of four possible different variants are considered. We selected third one for comparison, which is based on comparisons with \cite{Bruch:2004py}. In the NJL model, these numerators in the amplitude correspond to the following expression:

\begin{equation}
N_{V} = a_{V}\frac{4C_{V}}{g_{\phi}}I_{2}^{VKK}
\end{equation}

where the coefficient $C_{V}$, as shown in (\ref{gamma-in-meson}), is obtained from the quark loops in the transitions of the photon into the intermediate vector mesons and the numerical coefficients are $a_{\rho}=a_{\rho'}=1/2$, $a_{\omega}=a_{\omega'}=1/6$, $a_{\phi}=a_{\phi'}=1/3$.

\section{Conclusion}

The satisfactory agreement with the experimental data in the energy interval 1-1.6 GeV has been obtained in the extended NJL model for the $e^{+}e^{-} \rightarrow K^{+} K^{-}$ process. At the same time, in the region $> $1.6 GeV, the intermediate vector mesons $\rho(1700)$, $\omega(1650)$ significantly affect the final results. However, they are not taken into account in our version of the extended NJL model. So in this region we can not claim to satisfactory estimates with experiment. Once again, we emphasize that our results are obtained without any arbitrary parameters. The absolute values of the numerators of the Breit-Wigner propagators of the form factors that are qualitatively consistent with the results obtained in other phenomenological models are obtained.

\section*{Acknowledgments}
We are grateful to A. B. Arbuzov and S. B. Gerasimov for useful discussions.

\end{document}